\documentclass[11pt]{revtex4}
\usepackage{graphicx}
\usepackage{supertabular}
\usepackage{amsmath}
\usepackage{amsfonts}
\usepackage{amssymb}
\begin{document}
\title{Fermionization in an Arbitrary Number of Dimensions~\footnote{This is the talk published in 
the Proceedings to the $18^{th}$ Workshop "What Comes Beyond the Standard Models",
Bled, 11-19 of July, 2015.}}
\author{N.S. Manko\v c Bor\v stnik${}^1$ and H.B.F. Nielsen${}^2$\\
${}^1$University of Ljubljana,\\ Slovenia \\
${}^2$Niels Bohr Institute,\\
Denmark
%Copenhagen, DK-2100
}

%{\bf (This is a draft version by Holger 2013 december aiming at more general 
%allownce for extra dimensional space)}

%(This is not yet a paper but just a starting on the Proceedings by inserting the slides into a 
%paper)
\begin{abstract}
One purpose of this proceedings-contribution is to show that at least for free massless 
particles it is possible to construct an explicit boson theory which is exactly equivalent in 
terms of momenta and energy to a fermion theory. The fermions come as $2^{d/2-1}$ 
families and the to this whole system of fermions corresponding bosons come as a whole 
series of the Kalb-Ramond fields, one set of components for each number of indexes on 
the tensor fields.

Since Kalb-Ramond fields naturally (only) couple to the extended objects or branes, we 
suspect that inclusion of interaction into such for a bosonization prepared system - 
except for the lowest dimensions - without including branes or something like that is not likely 
to be possible.
 
The need for the families is easily seen just by using the theorem long ago put forward by 
Aratyn and one of us (H.B.F.N.), which says that to have the statistical mechanics of the 
fermion system and the boson system to match one needs to have the number of the field 
components in the ratio $\frac{2^{d-1}-1}{2^{d-1}}= \frac{\# bosons}{\# fermions}$, 
enforcing that the number of fermion components must be a multiple of $2^{d-1}$, where 
$d$ is the space-time dimension. This "explanation'' of the number of dimension is potentially 
useful for the explanation for the number of dimension put forward by one of us (S.N.M.B.) 
since long in the {\it spin-charge-family} theory, and leads like the latter to typically (a multiple of) 
$4$ families. 

And this is the second purpose for our work on the fermionization in an arbitrary number of 
dimensions   - namely to learn how "natural" is the inclusion of the families in the way the
{\it spin-charge-family} theory does.   
  \end{abstract}
%\maketitle
%



\maketitle
\section{Introduction}
\label{introduction}

This is the first draft to the paper, prepared so far only to appear in the Proceedings as the talk 
of one of the authors (H.B.F.N.. The contribution presents the 
main ideas for the  fermionization/bosonization procedure in any dimension $d$. More detailed
explanation will appear in the paper.
Although many things are not yet strictly proven, the 
 fermionization/bosonization seems, hopefully, to work in any dimensional space-time and also 
 in the presence of a weak background field. We hope, 
that the fermionization/bosonization procedure might help to better understand why nature has 
made of choice of spins, charges and families of fermions and of the corresponding gauge and 
scalar fields, observed in the low energy regime and why the {\it  spin-charge-family} 
theory~\cite{norma2014MatterAntimatter,scalarfieldsJMP2015} might be the right 
explanations for all the assumptions of the Standard Model, offering the next step beyond the
{\it standard model}. 

%Fermionization/Bosonization in Arbitrary 
%Dimensions for Free Particles, 
%Locality Predicts Four Families
%}}
%\author{\large \bf Norma Mankoc-Borstnik
%\footnote{ }\\ 
%\itshape{\large ${}^{1}$ Ljubljana,}
%%L.V.~Laperashvili
%%${}^{1}$ \footnote{laper@itep.ru}\,\,,
%H.B.~Nielsen${
%}^{2}$ \footnote{hbech@nbi.dk}\,\, 
%and A.~Tureanu ${}^{3}$
%\footnote{anca.tureanu@helsinki.fi}\\
%\itshape{\large ${}^{1}$ Ljubljana,}\\
%{\it National Research Center 
%"Kurchatov Institute", Moscow}\\
%\itshape{\large ${}^{2}$ The Niels Bohr %
%Institute, Copenhagen}\\
%}
%\itshape{\large ${}^{3}$ University of 
%Helsinki and
%Helsinki Institute of Physics, Helsinki}}
%\maketitle
%\end{frame}
%\begin{frame}
%\title{Main Subjects:}

%Fermionization/Bosonization in Arbitrary 
%Dimensions for Free Particles, 
%Locality Predicts Four Families
%}}
%\author{\large \bf Norma Mankoc-Borstnik
%\footnote{ }\\ 
%\itshape{\large ${}^{1}$ Ljubljana,}
%%L.V.~Laperashvili
%%${}^{1}$ \footnote{laper@itep.ru}\,\,,
%H.B.~Nielsen${
%}^{2}$ \footnote{hbech@nbi.dk}\,\, 
%and A.~Tureanu ${}^{3}$
%\footnote{anca.tureanu@helsinki.fi}\\
%\itshape{\large ${}^{1}$ Ljubljana,}\\
%{\it National Research Center 
%"Kurchatov Institute", Moscow}\\
%\itshape{\large ${}^{2}$ The Niels Bohr %
%Institute, Copenhagen}\\
%}
%\itshape{\large ${}^{3}$ University of 
%Helsinki and
%Helsinki Institute of Physics, Helsinki}}
%\maketitle
%\end{frame}
%\begin{frame}
%\title{Main Subjects:}

This talk demonstrates that:
\begin{itemize}
\item  Bosonization/fermionization is possible in an arbitrary number of dimensions (although
the fermions theories are non-local due to the anticommuting nature of fermions, while  bosons 
commute).
\item The number of degrees of freedom for fermions versus bosons obeys in our 
procedure in any dimension the Aratyn-Nielsen theorem~\cite{AH}.
%
%\item Fermions theories are non-local (if true fundametal fermions)
\item The number of families in four dimensional space-time is (a multiple of) four families.
\end{itemize}
%

%\begin{center}
To prove for massless fermions and bosons 
that the bosonization/fermionization is 
possible
in an arbitrary number of dimensions we 
use the Jacoby's triple product formula, 
presented
by Leonhard Euler in 1748~[20] 
%This result was proved by Leonhard Euler in 1748[20] 
which is  a special case of Glaisher's theorem 
\begin{equation}
\frac{1}{2} \prod_{n=0,1,2,...}(1+x^n)
= \prod_{m= 1,3,5,...}\frac{1}{1-x^m}\,.
\label{11ue}
\end{equation}
Let the reader notices that the product on the left hand side runs over 0 and all positive
integers, while on the right hand side it runs only over odd positive integers.
We shall connect the left hand side with the fermion degrees of freedom and the right hand side 
with the boson degrees of freedom, recognizing that for all positive numbers the number of 
partitions with odd parts equals the number of partitions with distinct parts [19]. 
% What follows must be explained better
%{\bf Odd parts and distinct parts}
%
Let us demonstrate this in a special case: \\
Among the 22 partitions of the number 8 there are 6 that contain only odd parts, namely\\
(7 + 1, 5 + 3, 5 + 1 + 1 + 1, 3 + 3 + 1 + 1, 3 + 1 + 1 + 1 + 1 + 1,\\  
1 + 1 + 1 + 1 + 1 + 1 + 1 + 1). This is to be connected with the boson (the right hand side 
of Eq.~\ref{11ue}) degrees of freedom.\\
 If we count partitions of 8, in which no number occurs more than once, that is with distinct 
parts,  we obtain again 6 such partitions, namely\\
(8, 7 + 1, 6 + 2, 5 + 3, 5 + 2 + 1, 4 + 3 + 1). This is to be correspondingly connected with
the fermion (the left hand side of Eq.~\ref{11ue}) degrees of freedom.\\

For every type of restricted partition there is a corresponding function for the number of 
partitions satisfying the given restriction. An important example is q(n), the number of partitions 
of n into distinct parts~\cite{euler1}.
The generating function for q(n), partitions into distinct parts, is given by~[22] 
% Check this equation
%
\begin{equation}
\label{euler1}
    \sum_{n=0}^\infty q(n)x^n = \prod_{k=1}^\infty (1+x^k) = 
\prod_{k=1}^\infty \frac {1}{1-x^{2k-1}}\, .
\end{equation}
The first few values of q(n) are (starting with q(0)=1):\\
(1, 1, 1, 2, 2, 3, 4, 5, 6, 8, 10, …)% (sequence A000009 in OEIS)).

% Correct below
%The second product can be written $\phi(x2) / \phi(x)$, where $\phi$ is the Euler's 
%function; 
The pentagonal number theorem can be applied % to this as well 
giving a recurrence for q~\cite{euler1}:
\begin{equation}
\label{euler1a}
    q(k) = a(k) + q (k-1) + q (k-2) - q (k-5) - q (k-7) + q (k-12) + q (k-15) - q (k-22)  - ...
\end{equation}
%
%Correct the equation below
where a(k) is $(−1)^{m}$, if $k = (3 m^2 - m)$ for some integer m, and is 0 otherwise.

\vspace{10mm}

\includegraphics[clip]{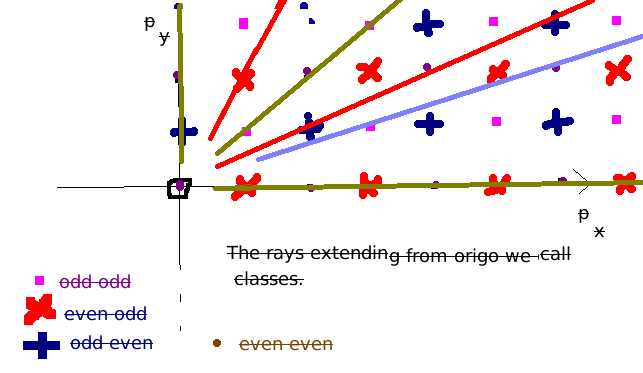}

\vspace{5mm}

\begin{center}
{\bf Bosonization Illustrating Formula}\\
\vspace{5mm}

Below the $d_{space}$ space dimensional version, $d=1+ d_{space}$, of Eq.~\ref{11ue} 
(for only a ``quadrant'') is presented.
\end{center}
\begin{eqnarray}
\frac{1}{2}\prod_{(m_1,m_2,...,m_{d_{space}})\in {\bf N}_0^{d_{space}}} 
(1+ x_1^{m_1}x_2^{m_2}\cdots x_{d_{space}}^{m_{d_{space}}})&=& \nonumber\\
=\prod_{(n_1,n_2, ..., n_{d_{space}})\in {\bf N}_0^{d_{space}}\hbox{but not all $n_i$'s 
even}}\frac{1}{1- x_1^{n_1}x_2^{n_2}\cdots x_{d_{space}}^{n_{d_{space}}}}&&\label{ue} 
\end{eqnarray}
\begin{eqnarray}
\frac{1}{2}\prod_{(m_1,m_2,...,m_{d_{space}})\in {\bf Z}^{d_{space}}}& & \, \nonumber\\
(1+ x_1^{m_1}x_2^{m_2}\cdots x_{d_{space}}^{m_{d_{space}}} 
z^{\sqrt{m_1^2 + m_2^2 + ...+ m_{d_{space}}^2}})&=& \, \nonumber\\
=\prod_{(n_1,n_2, ..., n_{d_{space}})\in {\bf Z}^{d_{space}}\hbox{but not all $n_i$'s 
even}}&& \, \nonumber\\
\frac{1}{1- x_1^{n_1}x_2^{n_2}\cdots x_{d_{space}}^{n_{d_{space}}}
z^{\sqrt{n_1^2 + n_2^2+...+n_{d_{space}}^2}} }\,.&&\label{g1} 
\end{eqnarray}

\vspace{10mm}

\begin{center}
{\bf The Idea for the Procedure for a Proof of the Existence of the 
Multidimensional/Bosonization Formula}
\end{center}

\vspace{5mm}

The idea for  generalizing  the bosonization/fermionization procedure from 
$d=(1+1)-$dimensional case, where the Euler's triple formula~\ref{11ue}  relates
the number of the Fock space basis vectors (states) for fermions and bosons, respectively, 
classified after the discretized momenta, where a fermion can have any positive or zero integer
momentum $n$  - the left hand side of Eq.~\ref{11ue} - while a boson can have any odd value
$m$ of the momentum. Both, the left and the right hand side manifest the same degrees of 
freedom. This can be seen if summing up on the fermionic side possibilities that $n$ of $x^n$ is 
a sum of all different integers,  (example 1: factor $2$ at $2x^4$ arises from the fact that 
$4=4+0=3+1$), while at the boson (right hand) side the factor at $x^m$  is the sum of all 
possibilities that $m$ is written as a sum of positive odd integers (example 1: factor $2$ at
$2 x^4$ arises from $4=3+1=1+1+1+1+1$).
  
% 23.01.2016 at 17:15 
We can generalize this  bosonization/fermionization procedure from $d=d_{space}+1=1+1$ 
to any dimensional space time $d = d_{space} +1$ as follows:\\

\begin{itemize}
\item{1.} We devide the whole system of all the discretized momentum vectors into 
"classes'' of proportional vectors 
(meaning in practice vectors deviating by 
a rational 
factor only), or rays (we might call them the rays of the module).
\item{2.} For each ``class'' the proof  is given by the 1+1 dimensional case which 
means by just using the formula by Euler and extending it to both positive and 
negative integers. 
\item{3.} For $d_{space} > 1$ it is meaningful to require that not only momentum components 
but also the energy are the same for fermions and bosons. 
\end{itemize}

\vspace{5mm}

% 23.01.2016 at 21.15
%
\begin{center}
{\bf Formulas of Bosonization as Products over Rays/Classes} 
\end{center}

We correspondingly  generalize Eq.~\ref{11ue} by generalizing $x$ by  a product of $x_{i}$, 
($i=1,\dots,d_{space}$) and $z$, the momentum $m$ by $m_i(c) \times m$ (or rather 
$m_i(c)*m$), the power of $z$ by  $\sqrt{\sum_i (m_{i}(c)\times m)^2}$ 
($\sqrt{\sum_i (m_{i}(c)*m)^2}$) on the fermion (the left hand) side, and correspondingly 
the momentum $n$ by $n_i(c) \times n$ ($n_i(c)*n$), the power of $z$ by
$\sqrt{\sum_i (n_{i}(c)\times n)^2}$ ($\sqrt{\sum_i (n_{i}(c)*n)^2}$) on the boson (the 
left hand) side, while the summation runs over the classes.
%\vspace{5mm}

%
\begin{eqnarray}
\label{g}
\prod_{c\in rays} \;\;\prod_{m \in {\bf Z}, m\ne 0}\hspace{3mm}&&\nonumber \\
\hspace{-2mm}(1+ x_1^{m_1(c)*m}x_2^{m_2(c)*m}
\cdots 
x_{d_{space}}^{m_{d_{space}}(c)*m} z^{\sqrt{m_1^2(c) + m_2^2(c) + ...+ 
m_{d_{space}}^2(c)}\;\;*|m|})&=&\nonumber \\
=\prod_{c \in rays}\;\;\prod_{n \hbox{ odd}}\hspace{3 mm}&& \nonumber \\
\hspace{-2 mm} \frac{1}{1- x_1^{n_1(c)*n}x_2^{n_2(c)*n}
\cdots x_{d_{space}}^{n_{d_{space}}(c)*n}
z^{\sqrt{n_1^2(c) + n_2^2(c)+...+n_{d_{space}}^2(c)}\;\;*|n|} }\,.&& 
\end{eqnarray}
where $c$ runs over the set $rays$ of the $d_{space}$-tuples of non-negative integers,
that cannot be written as such a tuple multiplied by an over all integer factor.  

\vspace{10mm}

\includegraphics[clip]{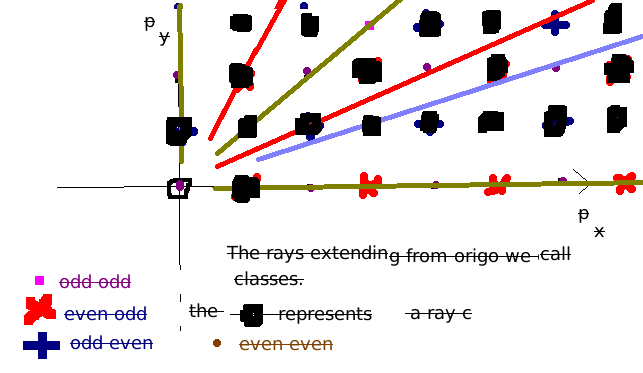}
%

%30.11.2015 at 11:00
\vspace{10mm}

\begin{center}
{\bf Splitting the Fock space into Cartesian Product Factors from Each Ray $c$} 
\end{center}

\vspace{5mm}

Denoting the Fock space for the theory - it be a boson or a fermion one - as ${\cal H}$
for the $d_{space}$-dimensional theory, and by ${\cal H}_c$ the Fock space for the - 
essentially 1 + 1 dimensional theory associated with the ray/(or class) $c$ describing  
the particles with momenta being an integer (though not 0) times the representative for 
$c$, namely $(m_1(c), m_2(c), ..., m_{d_{space}}(c))$ - it is suggested that we write 
the full Fock space as the product
\begin{equation}
{\cal H} = \otimes_{c \, \in RAYS} {\cal H}_c.
\label{firay}  
\end{equation}

\vspace{5mm}

\begin{center}
{\bf Introduction of Creation and Annihilation Operators}
\end{center}

\vspace{5mm}

We shall introduce for a boson interpretation of the Hilbert space  the Fock space
${\cal H}$:
\begin{eqnarray}
\hspace{-2mm} a (n_1, n_2, ..., n_{d_{space}}) & annihilates & \hbox{a boson with 
momentum}\; (n_1, n_2, ..., n_{d_{space}})\,,\nonumber \\
\hspace{-2 mm} a^{\dagger}(n_1, n_2, ..., n_{d_{space}}) & 
creates  & \hbox{ a boson with
momentum}  \; (n_1, n_2, ..., n_{d_{space}})\,, \nonumber
\end{eqnarray} 
where the integers can be any, except that they must {\em not} all $d_{space}$ ones 
be even.

Similarly for fermions:
\begin{eqnarray}
\hspace{-2mm} b (n_1, n_2, ..., n_{d_{space}}) & annihilates & \hbox{a fermion with 
momentum}\;  (n_1, n_2, ..., n_{d_{space}}) \,,\nonumber \\
\hspace{-2 mm} b^{\dagger}(n_1, n_2, ..., n_{d_{space}}) & 
creates  & \hbox{a fermion with momentum}\; (n_1, n_2, ..., n_{d_{space}})\,, \nonumber
\end{eqnarray}
where now the $n_i$ numbers can be any integers.

\vspace{5mm}

\begin{center}
{\bf Boson Operators Dividable into rays or classes $c$, also Fermion Operators Dividable into
 Rays or Classes $c$, Except for One Type}
\end{center}

\vspace{5mm}

% 30.11.2015 at 15:00

We can write any "not all even'' (discretized) momentum $(n_1, n_2,..., n_{d_{space}})$ as an 
{\em odd} integer $n$ times a representative for a class/ray $c$
\begin{eqnarray}
a(n_1, n_2, ..., n_{d_{space}})&=& a (n_1(c)*n, n_2(c)*n, ..., n_{d_{space}}(c)*n)\,,
\nonumber \\
a^{\dagger}(n_1, n_2, ..., n_{d_{space}})&=&
 a^{\dagger}(n_1(c)*n, n_2(c)*n, ..., n_{d_{space}}(c)*n)\,. %\nonumber
\end{eqnarray}
The boson momentum with a given even/odd combination for its momentum components 
(say $oe...o$) goes to a ray/class $c$ with the {\em same combination of even/odd-ness}.

Similarly one can proceed also for fermions with not all momentum components even; but
the fermion momenta that have all components even {\em get divided into rays/classes  
with different even/odd combinations}. There are no rays with the even combination 
$ee...e$, of course, because a tuple of only even numbers could be divided by 2.

\section{On Construction of Fermion Operators}
\label{thoughts}

We have made an important step  arriving at a model suggesting how it could be possible to 
match momenta and energies for a system with either fermions or bosons. To completely 
show the existence of fermionization (or looking the opposite way, bosonization) we should, 
however, write down the formula for how the fermion (boson) creation and annihilation 
operators are constructed in terms of the boson (fermion) operators, so that it can become 
clear (be proven) that the phase conventions and identification of the specific states with a 
given total momentum and energy for fermions can be identified with specific states for the
boson system. 

Such a construction is well known for 1+1 dimensions, where it looks like 
\begin{equation}
\psi_e(x) + i \psi_o(x) = \exp(i \phi_R(x))  
\label{pos}  
\end{equation}
in the "position'' representation, meaning that
\begin{eqnarray}
\label{position}
\psi_e(x) &=& \sum_{m \hbox{ even}}\exp(imx) \;b_e
(m) \, \nonumber\\
\psi_o(x) &=& \sum_{m \hbox{ odd}}\;\exp(imx) \;b_o
(m)  \, \nonumber\\
\phi_o(x) &=& \sum_{m \hbox{ odd}}\;\exp(imx)\;a_o(m)\,.
\end{eqnarray}
Let us think of the case of making the field operators in position space $\phi_e(x),\; 
\phi_o(x),\; \psi_o(x)$ Hermitean by assumming
\begin{eqnarray}
a_o(m) &=& a_o^{\dagger}(-m);\; \hbox{ for all  $m$ odd,} \nonumber\\
b_o(m) &=& b_o^{\dagger}(-m);\; \hbox{ for all  $m$ odd,} \nonumber\\
b_e(m) &=& b_e^{\dagger}(-m);\; \hbox{ for all  $m$ even}\,.
\end{eqnarray}
\subsection{The Problem of Extending to Higher Dimensions Even if we Have Bosonization Ray 
for Ray} 
\label{problem}

At first one might naively think that - since each of our rays (or classes) c functions as  the 1+1 
dimensional system and  we can write the whole fermion, as well as the whole boson, space
according to (\ref{firay}) - it would be trivial to obtain the bosonization for the whole system and 
thereby have achieved the bosonization in the arbitrary dimension, which is the major goal of this 
article. 

However, one should notice that constructing in a simple way a system composed from several 
independent subsystems such it is the whole system ${\cal H}$, composed from the subsystems 
${\cal H}_c$ (for $c\in rays$), one obtains {\em commutation} between operators acting solely 
inside one subsystem $c$, say, and operators acting solely inside another subsystem $c'$, say. 
But we want for the fermions the {\em anticommutation} relations rather than the commutation
ones, and thus some (little ?) trick is needed to achieve this anticommutation. 

First we shall show how this anticommutation can be achieved by means of an ordering of all the
rays $c \in rays$ by some ordering inequality being chosen between these rays: $>$. But this
is a very ugly procedure and we shall develop a slightly more general attempt in which we 
construct a phase $\delta(c,c')$ for each pair of rays $c$ and $c'$. Then we shall go on seeking 
to make the choice of this phase $\delta(c, c')$  in a continuous and more elegant way. Since
that shall turn out to be non-trivial,  we shall develop the ideas by first seeking for such a
construction of the phase for the odd dimensional space of $d=3$, meaning $d_{space}=2$, 
to learn the idea, although we are most keen on even space-time dimensions, such it is the 
experimentally observed number of space-time dimensions, $d=4$.

\subsection{The $>$ Ordered Rays Construction}
\label{ordered}

Let us suppose that we have a formal way of constructing the fermion creation and annihilation 
operators in terms of the boson operators. We do indeed have such a construction, since we can 
Fourier transform back and forth the construction in the position representation (\ref{pos}) and 
the 1+1 dimensional bosonization is so well understood. Since for the present problem the details 
of this 1+1 dimensional bosonization relations are not so important, we shall just assume that
we are able to deduce for each ray or class $c$ 
%can 
a series of fermion creation - 
$b^{\dagger}_{naive \; o}(m, c)$ and $b^{\dagger}_{naive \; e}(m, c)$ - and annihilation -
$b_{naive \; o}(m, c)$ and $b_{naive \; e}(m, c)$ - operators, that function well as fermion 
operators {\em inside the ray $c$}, so to speak. $o$ and $e$ denotes odd and even, 
respectively.
The only important think is that these operators {\em can be expressed in terms of the boson
%s 
annihilation and creation operators belonging to the same ray $c$}:
\begin{eqnarray}
b^{\dagger}_{naive \; o}(m, c)&=& 
b^{\dagger}_{naive \; o}(m, c; a_{o}(n, c)\,, \,\hbox{for } n \hbox{ odd}  )\\
b_{naive \; o}(m, c)&=& 
b_{naive \; o}(m, c; a_{o}(n, c) \,, \,\hbox{for } n \hbox{ odd}  )\\
b^{\dagger}_{naive \; e}(m, c)&=& 
b^{\dagger}_{naive \; e}(m, c; a_{o}(n, c) \,, \,\hbox{for } n \hbox{ odd}  )\\
b_{naive \; e}(m, c)&=& 
b_{naive \; e}(m, c; a_{o}(n, c)\,, \,\hbox{for } n \hbox{ odd})\,.
\end{eqnarray}
For these operators we know form the 1+1 dimensional bosonization that we can take them to 
obey the usual anticommutation rules {\em provided we keep to only one ray $c$}:
\begin{eqnarray}
\{ b^{\dagger}_{naive \; o}(m, c; a_{o}(n, c)\,, \,\hbox{for } n \hbox{ odd}  ),
 b^{\dagger}_{naive \; o}(p, c; a_{o}(n, c)\,, \,\hbox{for } n \hbox{ odd}  )\}_{-}&=& 
\delta_{n, -p}\,, \, \hbox{for $m,p$ both odd}\,,\nonumber \\
\{ b_{naive \; o}(m, c; a_{o}(n, c)\,, \,\hbox{for } n \hbox{ odd}  ),
 b^{\dagger}_{naive \; o}(p, c; a_{o}(n, c)\,, \,\hbox{for } n \hbox{ odd}  )\}_{-}&=& 
\delta_{n, p}\,, \, \hbox{for $m,p$ both odd}\,,\nonumber\\
\{ b_{naive \; o}(m, c; a_{o}(n, c)\,, \,\hbox{for } n \hbox{ odd}  ),
 b_{naive \; o}(p, c; a_{o}(n, c)\,, \,\hbox{for } n \hbox{ odd}  )\}_{-}&=& 
\delta_{n, -p}\,, \hbox{for $m,p$ both odd}\,.\nonumber 
\end{eqnarray} 
%
%02.12. 2015 at 11:30          
We have similar anticommutation rules for annihilation and creation operators if exchanging the
index $o$ (meaning odd) by the index $e$ (meaning even), but now we should take into account
that the fermion operators with zero momentum, i.e. $(m, p) =0$, are not constructed from a 
single ray $c$. Rather there are - referring to our little problem with the explicit factor 
$\frac{1}{2}$ in the state counting formulae - not enough degrees of freedom in the 1+1 
dimensional boson system to deliver a fermion operator with a zero momentum.
% in the $d$ dimensional case. 

We should therefore imagine that we do not have these zero momentum fermion operators 
attached to our rays either. This is actually good for our hopes of bosonizing in higher 
dimensions because the zero momentum fermion operators would have had to
%o 
be common for 
the infinitely many rays and we would have had too many candidates for the zero momentum
fermion mode. Now instead we totally miss the zero momentum creation and annihilation 
fermion operators for the many dimensional system. That is, however, not at all so bad as it
would have been to get an infinity of them, because we fundamentally can not expect to 
produce all fermion operators from boson ones because we cannot possibly build up a sector with 
an odd number of fermions from boson operators acting on say some vacuum with an even
number. Therefore one fermion operator must be missing. This becomes the zero momentum 
one and that is o.k..

Our real problem remains that these naive fermion operators taken for two different rays, $c$ 
and $c'\ne c$, will commute
\begin{eqnarray}
\label{commfer}
\{ b^{\dagger}_{naive \; o} (m, c; a_{o}(n, c) \,, \,\hbox{for } n \hbox{ odd} ),
    b^{\dagger}_{naive \; o}(p, c'; a_{o}(n, c')\,, \,\hbox{for } n \hbox{ odd}  )\}_{-}&=&0 \,, \,
 \hbox{for $m,p$ both odd}\nonumber\\
etc.\,. && 
\end{eqnarray}

We could define an $(-1)^F$-operator, where $F$ is the fermion number operator. 
It sounds at first very easy just to write
\begin{equation}
F_c = \sum_m b^{\dagger}_{naive\; o}(m,c) \;b_{naive \; o}(m,c) + 
\sum_m  b^{\dagger}_{naive\; e}(m,c)\;b_{naive \; e}(m,c),            
 \end{equation}
where the sums run over respectively the odd and the even positive values for $m$ for the 
$o$ and the $e$ components. But now this fermion number operator- as taken as a function of
the {\em naive} operators - ends necessarily up being an expression in purely boson operators 
(from the ray $c$), and thus it looks at first as being valid except when the expression 
$(-1)^{F_c}$, which we are interested in, is equal to $1$ on all states that can truly be 
constructed from boson operators. If it were indeed so, our idea of using $(-1)^{F_c}$ to 
construct the multidimensional fermion operators, would not be so good. However, there is a little
detail that we did not have enough bosonic degrees of freedom to construct the zero momentum 
fermion operator in 1+1 dimensions. Therefore we can not really include in the definition of the 
"fermion number operator  for the ray $c$'',  $F_c$,  the term coming from $m=0$. This term
would formally have been $b^{\dagger}_{naive \; e}(m=0, c) \; b_{naive \; e}(m=0,c)$, but 
we decided to leave it out. This then means that the fermion number operator, for which we 
decide to use $F_c$ as the number of fermions operator in the ray $c$ is {\em not the full 
fermion number operator for the corresponding 1+1 dimensional theory, but rather only for those 
fermions, that avoid the zero momentum state}. To require this avoidance of the zero 
momentum is actually very attractive for defining a fermion number operator {\em for the 
ray $c$} as far as the momentum states included in such a ray really must exclude the zero 
momentum state in a similar way as a ray in a vector space is determined from the set 
of vectors in the ray not being zero.

But this precise definition avoiding the zero-momentum fermion operator contribution to the 
fermion number operator $F_c$ leads to the avoidance of the just mentioned problem that
this fermion number $F_c$ looked as always having to be even when constructed in terms of
boson operators.
%02.12.2015 at 14:15 

Now there should namely be enough boson degrees of freedom that one should be able to 
construct by boson operators all the different possible combinations for fermion states being 
filled or unfilled (still not the zero momentum included). Thus one does by pure bosons construct 
both - the even $F_c$  and the odd $F_c$ - states and thus the $F_c$ with the zero momentum 
fermion state not counted can indeed be a function of the boson operators and can {\em take 
on both even and odd values} for momentum, depending on the boson system state.  So,
we can have - using this leaving out the zero momentum fermion state in the rays -  
an operator 
\begin{equation}
F_c = F_{naive \; c}(a_e(n) \,,\,\hbox{ for $n$ odd})
\end{equation}   
%
%24.01.2016 at 12:15 
The operator $(-1)^{F_c}$ for each ray $c$ counts if the number of fermions in the 1+1 
dimensional system is even, then  $(-1)^{F_c}=1$, or odd, then  $(-1)^{F_c}=-1$. We 
construct the following improved fermion operator (annihilation or creation),
\begin{eqnarray}
b_{e}(m,c) &=& b_{naive \: e}(m,c) \prod_{c'<c}
(-1)^{F_{c'}}.
  \end{eqnarray}
The inclusion of this extra operator factor helps to convert the commutation relations between the 
fermion annihilation and creation operators for different rays into anticommutation  relations, 
as it can easily be seen
\begin{eqnarray}
&&b_{e}\,(m, c)\; b_e\,(p,c') =
      b_{naive \; e} \,(m, c) \cdot %\left (
 \prod_{c'' <c} \,(-1)^{F_{c''}} %\right ) 
\;\; b_{naive \; e}\, (p,c') \cdot %\left ( 
\prod_{c'''<c'} \,(-1)^{F_{c'''}} % \right ) 
= \nonumber\\
&& b_{naive \; e}\, (m, c) \cdot % \left ( 
%\prod_{c'' <c}\, (-1)^{F_{c''}} %\right ) 
%b_{naive \; e} \,(p,c') \left ( \prod_{c'''<c'}
%(-1)^{F_{c'''}} \right )= \\
%= b_{naive \; e}(m, c)
%\left ( 
\prod_{c'\le c'' <c} \,(-1)^{F_{c''}} %\right )  
\;\;b_{naive \; e}\, (p,c')\,% ,\,\hbox{for $c>c'$} 
= \nonumber\\
&& - b_{naive \; e}\, (m, c)\;\; b_{naive \; e} \,(p,c') \cdot %\left ( 
\prod_{c'<c''<c}\,(-1)^{F_{c''}} %\right )&=&
=\nonumber\\
&& - b_{naive \; e}\,(p, c')\;\; b_{naive \; e}\,(m,c) \cdot%\left (
 \prod_{c'<c''<c} (-1)^{F_{c''}} % \right )\\
= \nonumber\\
&& - b_{naive \; e}\,(p, c') \cdot %\left ( 
 \prod_{c'''<c'} \,(-1)^{F_{c'''}} %\right ) *  
\;\;b_{naive \; e}\,(m,c) \cdot % \left (
 \prod_{c''<c} (-1)^{F_{c''}} %\right )& =&
\nonumber\\
&& =- b_{e}(p, c')\; b_e(m,c )\,,\, \hbox{ still for $c>c'$}\,. 
\end{eqnarray}
Thus we deduced, for $c>c'$ in our in fact at first just chosen ordering of $<$, that the fermion 
operators %constructed 
do anticommute. It is not difficult to show similarly also in the case $c'> c$, that the fermion 
operators anticommute. The crux of the matter is that when e.g. $c'>c$ there is the factor 
$(-1)^{F_{c}}$ contained in the product $\prod_{c''< c'}(-1)^{F_{c''}}$, which is attached to 
$b_{naive \; o}\,(m,c')$ in order to correct it into $b_{o}\,(m,c')$, while there is no analogous
factor $(-1)^{F_{c'}}$ contained in the factor $\prod_{c'' <c}(-1)^{F_{c''}}$ 
attached at $b_{naive  \; o}\,(m, c)$ in order to bring it % to make it 
into  $b_{o}\,(m, c)$. In this way one gets just the one extra minus sign in the product of the 
fermion operators that makes them anticommute.

\subsection{Slight Generalization to have a Phase Factor}
\label{slightgen}

It is not difficult to see that the idea of using such an ordering $<$ could be slightly generalized
to have instead of the factors only minus or plus phase factors of the form $\exp(\delta(c,c'))$
% meaning that one constructs the ansatz for the true fermion creation operator say
%
\begin{eqnarray}
b^{\dagger}_{e}\,(m,c)&=& b^{\dagger}_{naive \; e}\,(m,c) \prod_{c'\ne c, 
\hbox{ but }c'\in rays } e^{(i\delta(c,c') F_{c'})}\,.
\label{phansatz}
\end{eqnarray}
It is also not difficult to see that, in order to obtain the anticommutation relations instead of the 
commutation ones (which we have for $b^{\dagger}_{naive \; e}\,(m, c)$), the phases must 
obey the rule 
\begin{equation}
\delta(c,c') - \delta(c',c) = \pi (mod \; 2\pi)\,.
\label{picon}
\end{equation} 
We may in fact seek to evaluate the product  of two fermion creation operators  with the 
ansatz~(\ref{phansatz})
\begin{eqnarray}
&& b^{\dagger}_{e}\,(m,c) \;\;b^{\dagger}_{e}\,(m',c')=\nonumber \\ 
&& b^{\dagger}_{naive \; e}\,(m,c) \prod_{c''\ne c, \hbox{ but } c''\in rays} 
e^{i\delta(c,c'')\,F_{c''}} \;\; b^{\dagger}_{naive \; e}\,(m',c') \prod_{c'''\ne c', 
\hbox{ but }c'''\in rays} e^{i\delta(c',c''') F_{c'''}}=\nonumber \\ 
&& b^{\dagger}_{naive \; e}\,(m,c)\;e^{i\delta(c,c')} \prod_{c''\ne c \hbox{ nor } c', 
\hbox{ but }c''\in rays} e^{i(\delta(c,c'')+\delta(c',c'')) F_{c''}}\;\; 
b^{\dagger}_{naive \; e}\,(m',c') =\nonumber \\
&& b^{\dagger}_{naive \; e}\,(m,c) \;e^{i\delta(c,c')F_{c'}}\;\; 
b^{\dagger}_{naive \; e}\,(m',c')\;e^{i\delta(c',c)F_c}  \prod_{c''\ne c \hbox{ nor } c', 
\hbox{ but }c''\in rays} e^{i(\delta(c,c'') +\delta(c',c'')) F_{c''}}=\nonumber \\
&& e^{i(\delta(c,c') - \delta(c',c))}\;b^{\dagger}_{naive \; e}\,(m',c')\;
e^{i\delta(c',c)F_{c}} \;\; b^{\dagger}_{naive \; e}\,(m,c)\; e^{i\delta(c,c')F_{c'}} 
\nonumber \\
&&\cdot \prod_{c''\ne c \hbox{ nor } c', \hbox{ but } c''\in rays } e^{i(\delta(c,c'') +\delta(c',c'')}
F_{c''})=e^{i(\delta(c,c') - \delta(c',c))}\;b^{\dagger}_{e}\,(m',c')\;\; b^{\dagger}_{e}\,(m,c)
=\nonumber \\
&& - b^{\dagger}_{e}(m',c')\;\; b^{\dagger}_{e}(m,c)\,,
\label{phansatzuse}
\end{eqnarray}
where in the last step we used~(\ref{picon}). Thus we see that in this way we can get - really
in infinitely many ways - some algebraicly defined fermion operators that do indeed anticommute 
as they should. But it should be had in mind that both these procedures, by choosing  $\delta(c,c')$ 
and the forgoing proposal with the ordering $<$, are a priori discontinuous and arbitrary. 

We expect, however, that the latter method with $\delta(c,c')$ can be lead to a smooth 
and attractive scheme in the case of $d_{space}=2$ or equivalently $d=3$.

\subsection{Exercise with Next to Simplest Case $d_{space} =2$}
\label{exercise}

In the case of $d_{space}=2$ we can say that the set of our $rays$ form a kind of a 
set of  "rational angles'' in the sense that each ray specifies an angle modulo $\pi$ (rather than
modulo $2\pi$ as for an oriented arrow it would specify), but that one only obtains those angles 
which rationalize tangenses. But the fact that they are after all implemented as angles - although 
only modulo $\pi$, means that they are at least locally ordered as numbers along a real or rather 
rational axis. So apart from troubles at the end and beginning we have an ordering and we could 
attempt to use it even for the implementation of the ordered set of rays method by proposing
a "nice'' $<$ ordering. However, we think we get a better chance by using the $\delta(c,c')$ 
method in this $d=3$ and thus $d_{space}=2$ case. 

We have to think about what topological properties we shall and can achieve for the function 
$\delta(c,c')$ depending on a pair of rays $c$ and $c'$.   

Since the classes or rays are "a kind of rational'' directions, though without orientation, the
topological space of the rays is like the sphere $S^{d_{space}-1}$ with opposite points 
identified. This topological space obtained by the identification of the opposite point on the
 $S^{d_{space}-1}$ sphere is actually topologically identical to the projective space of 
$d_{space} -1$ dimensions. For the case $d=3$ or $d_{space}=2$ the topological space
rays thus become simply the projective line (using real numbers), but that is topologically 
just the $S^1$ circle. Had this topological space been naturally orderable we could have 
used the ordering as the $<$ above. However, it is a circle $S^1$ and not a simple line with plus 
and minus infinity; the infinities have so to speak been identified to only one point in the 
projective line. This means that using the method to define the fermion fields/operators 
by means of $<$-method would be very non-elegant, and would probably violate almost 
everything wanted.

Let us now think about a slight generalization by using the $\delta(c,c')$. We need to 
make a choice of a function $\delta(\cdot,\cdot)$ defined on the cross product of two 
projective spaces of dimension $d_{space}-1$ each. Since it shall obey the 
condition~(\ref{picon}), it cannot at all be a smooth or continuous function at the points 
where $c=c'$.
% So we must give that up by necessity. 
Let us, for a while, take care that this method works well for $d=3$ only.

In this $d=3$ case the cross product of the two projective lines becomes topologically simply 
a two-dimensional torus. So we face topologically to define $\delta(c,c')$ on a two-dimensional
torus. However, we are forced to give up having continuity along the "diagonal'' - meaning the 
set of points on this torus with $c=c'$ - and it is thus rather a $\delta(c,c')$ defined as a 
continuous function on the torus minus its "diagonal'', which we must choose/find. 

This two dimensional torus minus its "diagonal'' is rather like a belt. I.e., it is topologically like
the outer surface of a finite piece of a tube. It has two separate edges, each being topologically 
an $S^1$ circle, namely two images of the "diagonal'' seen from the two sides. In between
there is then the two-dimensional bulk area of the topological shape of the surface of the finite 
piece of a tube. It is inside this bulk region that we shall attempt to
construct $\delta(c,c')$ to be smooth and "nice''.
%({\bf the next sentence is probably 
% wrong:})
Choosing
\begin{equation}
\delta(c,c') = 2\hbox{``clock average angle''}(c,c')
\label{prop} 
\end{equation}
might be a good choice. 
Here  $\hbox{``clock average angle''}(c,c')$  is the angle that the unoriented direction  of 
slightly complicatedly defined angle-dividing line dividing the angle between the unoriented 
directions $c$ and $c'$   forms with some coordinate axis (in momentum space). 
The precise way of defining this $\hbox{"clock average angle''}(c,c')$ is %attempted 
illustrated on  the figure.
\begin{center}
\includegraphics{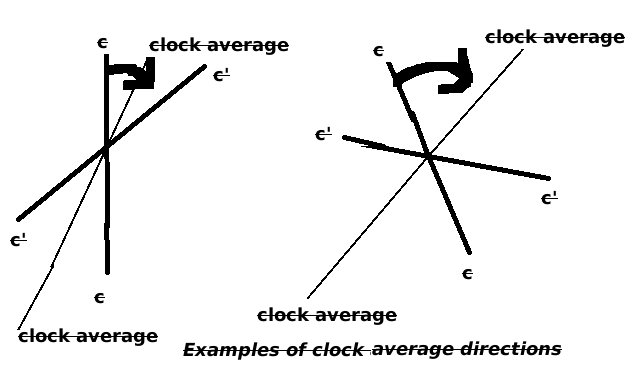}
\end{center}
It consists in the following (let us remind the reader that we are still  in the 
$d=3; d_{space}=2$ case):
\begin{itemize}
\item{a.} We introduce a "clockwise rotation orientation'' in the spatial momentum plane.
\item{b.} We draw a circle arrow from one of the two "ends'' (half lines) of which the line $c$
(the ray $c$ is basically just a line) in this clockwise direction, and note the angle between this 
end of $c$ and the first "end'' (=half  line) of $c'$ (met in the following the circular arrow), which
measures less than $180^0$.
\item{c.} We draw a line, that divide this under b. noted angular region into halves. This line
through the (momentum space) origo is denoted "clock average" (as marked on the figure).
\item{d.} Such an unoriented line as  the "clock average'' defines relative to a coordinate system
in spatial momentum space an angle-value modulo $\pi$. We call this angle-value 
$\hbox{``clock average angle''}(c,c')$ and it is as just said defined modulo $\pi$ (but only 
modulo $\pi$, because the line ``clock average'' is unoriented).
\item{e.} Multiplying this angle - $\hbox{``clock average angle''}(c,c')$ - by 2 its ambiguity to 
be only defined modulo $\pi$ becomes instead an ambiguity modulo $2\pi$. Thus our proposed 
expression~(\ref{prop}) for $\delta(c,c')$ is defined modulo $2\pi$, and that is what we need, 
since in our construction we exponentiate $\delta(c, c')$ after multiplication by $i$ and an 
operator $F_{c'}$ that has only integer eigenvalues. Thus the expression, which we use,  
$\exp(i\delta(c,c')F_{c'}) $ becomes well defined even though $2\hbox{"clock average angle''}\,
(c,c')$ makes sense only modulo $2\pi$.    
\end{itemize}

{\bf Let us see whether this proposal is indeed a good one.}
To see that our proposal~(\ref{prop}) is a good one we must first of all check that it 
obeys~(\ref{picon}). That is we must see what happens to the expression when we permute 
the two independent variables $c$ and $c'$. Since by definition the circular arrow constructed in
step b. goes out from the $c$-line,
% meaning
the first of the two arguments in $\delta(c,c')$, we must draw  this circle-arrow after the 
permutation from $c'$ instead. Therefore the half-angle noted under point b. above will after the 
permutation differ from the one before the permutation. This means that the line (through the 
origo) "clock average'' gets after the permutation perpendicular to its direction before the 
permutation of $c$ and $c'$.
%t
Therefore $\hbox{"clock average angle''}(c',c)
=\hbox{``clock average angle''}(c,c')+\pi/2 \hbox{(mod $\pi$)}$, which means that this angle 
gets shifted modulo $\pi$ with $\pi/2$. After the multiplication by  $2$ (point e.) it means that 
$\delta(c',c)= \delta(c,c') +\pi (mod \;2\pi)$, which is just~(\ref{picon}). Thus we got indeed 
by proposal~(\ref{prop}) the condition~(\ref{picon}) fulfilled. 

We can now remark that quite obviously our proposal~(\ref{prop}) is continuous as function 
of the directions $c$ and $c'$ {\em except where $c$ and $c'$ just coincide - what means that
it is zero (mod $\pi$) angle between them}. 
% 02.12.2015 at 1o:47

Let us note that had we not chosen the clock-wise rule, but instead taken, say, the smallest 
angle between $c$ and $c'$ and just found the halfening line between those "ends'', we would
have got a discontinuity when $c$ and $c'$ were perpendicular to each other. But by our 
precise choice we avoided that singularity.
%
%({\bf I think the following 5 lines may now be obsolete:})
%Let us understand the need for the singularity at the "diagonal''.
(For a point close to the diagonal the two arguments, $c$ and $c'$, are approximately the same 
ray. Permuting them will for a continuous function $ \delta(c,c')$ make almost no difference, and 
thus it cannot possibly change by $\pi$, while crossing the "diagonal'' the function $\delta$ would
ask to jump by $\pi$.) 

\section{A Guess for Arbitrary Dimension}
\label{guess}
%

%Very far out yet in completing I dream that we shall end up with a solution that 

We propose the generalization of Eq.~(\ref{pos}) to an arbitrary dimension, due  to our 
experience with the Clifford objects  (apart from some modifications due to
whether we choose Weyl or Majorana fermions for family or for geometrical components), by
using the relation
\begin{eqnarray}
(\psi + \psi_{m} \gamma^{m} + \psi_{mn}\gamma^{m}\gamma^{n}+...
+\psi_{1235...d} \,\Gamma^{(d-1)}&=& e^{ \phi_{m} \gamma^{m} + \phi_{mn}
\gamma^{m}\gamma^{n}+...
+\phi_{1235...d}\,\Gamma^{(d-1)}} \,. %``family column''.
\end{eqnarray}
% 
%  And then we should use Baker Hausdorf to show the correspondance.

%
\section{Outlook on Supporting the {\it Spin-Charge-Family} 
theory~\cite{norma2014MatterAntimatter,scalarfieldsJMP2015}}
\label{outlook}
%

%02.12.2015 at 14:20

We started with massless noninteracting bosons or fermions. But we want to work with the
interacting fields. There are many Kalb-Ramond fields appearing in our type of fermionizable 
boson model in higher dimensions and correspondingly it is not easy to see how to make an
interacting theory. 

There are many ways to come from noninteracting bosonisable (fermionizable) fermion
(boson) fields, which might lead to the fermion fields interacting with the boson fields as it is
in the {\it spin-charge-family} theory. 

But on the level of our fermionizable (bosonizable) boson (fermion) model with many
Kalb-Ramond fields we must keep in mind that the conserved charges in the Kalb-Ramond 
theories are vectorial and thus one gets very many vectorial conserved quantities. This 
makes scattering processes (unless all the scattering particles are without these vectorial 
charges) very non-trivial.
% It therefore does not look promising to introduce into our bosonization interactions just keeping
% particles unless the interaction totally avoids getting Kalb-Ramond charged particles into it. 

One chance would be to let either fermion or boson fields to interact with gravity.
% Comment:
% I still see that $f^{\alpha}{}_{c} {\cal S}^{ab}\omega_{ab \alpha}$, which is a pure boson
% field, applying on boson fields only,  goes into $f^{\alpha}{}_{c} S^{ab}\omega_{ab \alpha}$
% $+f^{\alpha}{}_{c} \tilde{S}^{ab} \tilde{\omega}_{ab \alpha}$, where either $S^{ab}$ or
% $\tilde{S}^{ab}$ apply on fermion fields, although I do not see yet how should the evolution
% go to come from our fermionization/bosonisation stage with Kalb-Ramond fields present to 
% the final stage of the {\it spin-charge-family} theory.
Crudely speaking gravity couples to energy and momentum, and since in the free bosonization 
procedure we have at least sought to get the total d-momentum be the same in the
corresponding states of fermions and  bosons there might be a chance that we fermionize a 
theory with both - the bosons of  the Kalb-Ramond type {\em and} gravity through the vierbein
formulation - and correspondingly obtain a theory with both fermions {\em and} bosons, the 
later would be the gravity degrees of freedom. This might lead to exactly the
 theory~\cite{scalarfieldsJMP2015,norma2014MatterAntimatter} that one of 
us (N.S.M.B.) has postulated as the true model for Nature beyond the standard model (the
{\it spin-charge-family} theory). 

Since our scheme a priori looks to require the Majorana fermions to have real fields like the
bosons - at least in the simplest version - we %shall presumably 
only expect to get chiral fermions in those dimensions wherein Majorana fermions can 
simultaneously be Weyl (which means chiral)~\cite{hnmajorana} 
as in d=2,6,10,14,... It is therefore even a slight support for the {\it spin-charge-family} theory 
that its phenomenologically favoured dimension is just 13+1 =14. 

One should for appreciating this idea of adding gravity without fermionizing it have in mind that 
one {\em does not have to fermionize (bosonize) all degrees of freedom}, but rather can - if one wishes - 
decide to fermionize (bosonize) 
some degrees of freedom but not all. Especially, if the motivation were 
to make all fermions from bosons because one claims that fermions are not properly local and
should not be allowed to exist, then of course it is enough that we start with a purely boson
theory as the fundamental one - and then we better only fermionize a part of bosons unless we
could identify a purely fermionic theory with nature. But of course there seemingly are bosons in 
nature and we thus must end phenomenologically with a theory with {\em both } bosons and 
fermions. 

Starting  from fundamental bosons only that is only achievable by only a {\em partial
fermionization}.

\begin{center}
{\bf Hope for the Progress}
\end{center}
The hope is, which is evidently from what we have proposed in this contribution,  that we shall 
construct 
formulas for the higher dimensional cases by generalizing the formulas we already have for the 
one dimensional case, generalizing as well the "classes" to higher dimensions. In the spirit of 
seeking to identify the fields characterized by their "odd/even'' indices with spin components, we
hope to derive from the %counting
bosonization formula a scheme formally stating the relation between the boson and the fermion 
second quantized fields,  
%Ther would we already can see be 
$2^{d_{space}}-1$ boson field components, while there will be $2^{d_{space}}$ fermion 
components.

\section{Outlook on the Connection to the {\it Spin-Charge-Family} Theory}
\label{outlook}

Let us try to clarify how the here discussed fermionization procedure is supposed to be, so to 
speak, the root for a theory beyond the {\it spin-charge-family} theory of Norma Susana Manko\v c
Bor\v stnik~\cite{norma2014MatterAntimatter,scalarfieldsJMP2015}  (and her collaborators),
 The (one of) way(s) we see as a very promising hope that one could
% to some extend 
justify this {\it spin-charge-family} theory by the hoped  fermionization is as follows:

{\em We build up a model with only bosons as the fundamental theory  in say - 13 +1 
dimensions - in the sense that this 13 +1 dimensional purely bosonic theory with a series of the
Kalb-Ramond fields {\em and} with usual 13+1 dimensional gravity should be the fundamental 
choice of nature} (not necessarily starting in d=13+1). Then this theory should be {\em partly}
fermionized in the sense that only the series of Kalb Ramond fields get fermionized, but not the 
gravity (bosonic) degres of freedom. The latter remain gravitational degrees of freedom
hopefully now functioning as gravity for the fermions that came out of the fermionization. 
%The  set up of this picture in which 
The {\it spin-charge-family} theory will show up out of the Kalb-Ramond components.
% with time indices on which we hope to come back to in a later article, we shall now describe 
%a bit closer:  
% 
\begin{enumerate}
\item The first assumption of our new scheme, which might be the pre-scheme of the 
{\it spin-charge-family} theory, is that {\em fermions a priori do not obey proper locality}. 
The accusation towards all the theories with fermions which are fundamental fermions rather
than fermionized bosons is that the axiom of locality in a quantum field theory is for the fermions
\begin{align}
\{ \psi_{\alpha}(x), \psi_{\beta}(y)\}_+ =0\,, \;\; \hbox{for the space like separation of $y$ and 
$x$,}
\end{align}
while true physical locality should have been a {\em commutation rule} like the one obeyed 
by the boson fields
\begin{align}
\{\phi(x) , \phi(y)\}_{-} =0\,, \;\;\hbox{for the  space like separation of $x$ and $y$.}
\end{align}
True locality % should 
means, one would think, that each little region in space is approximately a completely 
separate system that only interacts very indirectly with a far away different little region. If so, 
%then % it should mean that 
the physical operators describing the situation in one little region should commute with those 
describing the situation in a different little region, and not anticommute as the fermion fields do. 
One might like to assume  that only products of an even number of fermion fields are 
considered as proper operators describing the little system region, 
% With such a philosophy of only using pairs or even number products of fermion fields we could 
what satisfies the requirement of getting commutation relations between the field variables 
describing the situation in different regions. But such an assumption must be justified as
%we would presumably claim that also whether there is an even or an odd number of fermions
% in some little  region should be considered in principle 
a physical assumption, discussing seriously also  
%we should also take seriously the 
odd products of fermion fields. 

The point of view we suggest here is that we admit that we cannot have fermions at all in a 
truly local way! This then means that the fundamental physics should be a model {\em without
fermions} so that all fermions come from bosons that become fermionized.
\item  Since it is not easy to find so terribly many systems of bosons that can be fermionized, 
and thus if one finds some way of fermionizing, then this way is presumably already likely to 
be almost the only one possible. 

At least we expect that the fermionization of a boson system of fields can only be made 
provided the number of fermions and the number of bosons agree with the 
theorem which one of us (H.B.F.N.) and Aratyn~\cite{AH} put forward many years ago. 

For massless free fermions on the one side and massless free bosons on the other side
we obtained  that the number of components for the bosons and the fermions counted 
in the same way with respect to the fields being real or complex, should be in the ratio
\begin{equation}
\label{AtH}
\frac{\# \hbox{fermion components}}
{\# \hbox{boson components}} = 
\frac{2^{d_{space}}}{2^{d_{space}}-1}\,, 
\end{equation}
where the dimension is $d=d_{space}+1$, or the spatial dimension is $d_{space}$.

 The number of components - at least the number of real counted components - must 
of course 
be positive  integer or zero. Thus the minimal number of fermion components must be 
$2^{d_{space}}$, while the number of boson components must be $2^{d_{space}}-1$ or the 
numbers must be an integer multiplum of these numbers. 

Alone this theorem of ours~\cite{AH} makes appreciable restriction for when 
bosonization/fermionization is at all possible.
% SAID under 2.
%\item We may take this suggestion %that bosonization is not possible for so very many 
%combinations of fermion and boson fields, but rather a somewhat seldomh appening, to 
%suggest that if we find just one possiblity it is already likely that it is ``the right one'' in 
% the sense that if Nature needs to have bosonization, then it better take the one we can find.
%There are namely very likely not so many.
%
%04.12.2015 at 12:00
%
\item We  are suggesting here the starting point with the bosonic degrees of freedom only, 
consisting  of  "series of the Kalb-Ramond fields, all the chain through, except for one 
(pseudo)scalar". By this we mean  that we have as the bosons a series of separate fields 
$A_{mn \cdots r}$ with all the values of the number of indexes, antisymmetric with 
respect to all their indexes. 

There is a simple way in which one could get the number $2^{d_{space}}-1 $ of 
%components needed for the number of
boson components, if we arrange to have - by some gauge choice - only spatial values of the
indexes $m$, $n$,... on the A-fields, removing the $A$ field with zero indexes. 
%
% The reader should have in mind that a whole series of such fields with spatial components 
% would be specified by which subset of the spatial set of coordinates is represented as index. 
The number of components become equal to the number of subsets of $d_{space}$ letters, 
which is $2^{d_{space}}$. Removing %one by saying, but we do not include the 
pure scalar, we get this number %$2^{d_{space}}$ get reduced to 
$2^{d_{space}}-1$, as we want for the theorem of\cite{AH}. 
\item %If we go for this series of
From the $2^{d_{space}}-1$ bosons represented by the Kalb-Ramond fields (with the scalar 
removed), then according to the theorem ~\cite{AH} there must be  the  $2^{d_{space}} $ 
components of fermion fields. This means for the Weyl spinor representation of fermion fields
in even $d=d_{space} +1$, with $2^{d/2-1}$ members that there are $2 \times$ $2^{d/2-1}$
real fermion fields. To get $2^{d_{space}} $  real Weyl spinor representation fermion fields 
there must be  $\frac{2^{d_{space}}}{2^{(d_{space}+1)/2}} = $  $2^{d/2 -1}=$ 
$2^{d_{space}/2-1/2} $ families.
\item  From the bosonization requirement we obviously get out that there must exist an 
even number of families as it also comes out from the {\it spin-charge-family} theory of 
one of us (N.S.M.B.)~\cite{norma2014MatterAntimatter,scalarfieldsJMP2015}.
\item  But now there is a correction due to the components of the Kalb-Ramond fields with 
time indices, that is with $0$. This gives
very interesting corrections as we may postpone till later.          
\end{enumerate}

\subsection{A Hope for that the Gravity Interaction Can Be Added}

There is an interesting hope for that actually our at first free bosons being fermionized to 
free fermions could be generalized %a bit so that at least they could be allowed 
to have an universal coupling to a gravitational  field - the bosonic field, which we do not
fermionize, keeping it as gravitation,  % functioning also as the 
interactiing with the fermions - so that we finally arrive at a theory with several families of 
fermions and gravity.

In this contribution we wrote down a formula for  counting the number of states for the 
fermion and 
the boson systems having the same number of Fock states with given momentum and 
energy for the free massless case of our bosonization/fermionization procedure.

We used  in reality an infrared cut off that meant that we in fact considered a torus world 
with for different components different periodicity conditions: Some components of fields 
had antiperiodicity while the others had periodicity property along various coordinate directions.

We shall note now that we could consider these momentum eigenstates for the single particles 
with given periodicity restriction as {\em topologically} specified in the following sense:

The wave functions for the momentum eigenstates are as is well known all along taking on
only pure phase factor values, i.e. they obey $|\phi(x)|=1$ all along. The number of turns 
around zero, which they perform when one goes around the torus along the different 
coordinates, is an integer (or a half integer depending on the boundary condition).
We can consider this number of turns going around the torus in different ways (along different 
coordinates) a topological quantity in the sense that it as an integer cannot change under a 
small deformation. 

Our main idea is at this point that we in this way can introduce at least a not too strong
gravitational field and still have single particle solutions to the equations of motion characterized
by the same system of (topological) quantum numbers. 

That should suggest that we have the same set up for making the in this work studied 
bosonization in a not too strong gravitational field as in the free case. We namely should be 
able to classify the single particle states as functions of the space-time variables $x$ on the by 
gravitational fields deformed torus (torus due to infra red cut off) according to a topological 
classification in terms of the number of times the wave function encircles the value zero in 
the complex plane when the one follows a closed curve, following, say,  the coordinates of the 
deformed torus. For the massless theory we have scale-invariance for the matter fields - the 
series of the Kalb-Ramond fields or the fermions - so, as long as we consider the gravitational field 
as a background field,  i.e. we ignore the dynamics of the gravitational field itsef - we can scale
up the momenta of the single particles by just letting the phase of an eigen-solution be scaled
up by a factor. Only the periodicity conditions will enforce such scalings to be by integer factors, 
just as they must be also in the free flat case.

So we argue that with a background gravitational field, that is with a not too strong field, we 
have a possible description in terms of a discretized enumeration quite in the correspondence 
with the one for the flat case. 

Remembering that we obtained the bosonization w.r.t. state counting in fact class $c$ for class,
meaning that the momentum eigenstates in the classes corresponding to rays went separately 
from boson to fermion or oppositely, we may have given arguments at least suggesting 
that a corresponding bosonization correspondence as the one in the free flat case also applies
to the case with some (may be not too large though) gravitational field as a background field.

This may require further study but we take it that there is at least a hope for that the
 bosonization/fermionization procedure {\em can also be performed in a background 
gravitational field}. 

Since we now with our expansions in power series seek to guarantee that we shall make the 
bosonization or fermionizations just in such a way that the d-momentum will be the same for
the fermion configuration and the boson configuration corresponding to each other, we might 
hope that we could formulate the exact correspondence and the interpretations in terms of 
fields with spin indexes so that indeed the momentum densities would be the same for the 
fermions and for the corresponding bosons. If we succeed in that then the action on the 
gravitational fields which only feel the matter via the energy momentum tensor $T_{\mu\nu}$
would be the same for the bosons and the fermions in the corresponding states. In that 
case the development of the gravitational fields would be the same for the corresponding 
fermion and boson configurations. Thus the bosonization/fermionization  procedure would 
truly have been made also in the with gravity interacting models. Just  the gravity field itself 
should {\em not be fermionized}.

\end{document}